\documentclass[%
reprint,
superscriptaddress,
amsmath,amssymb,
aps,
prl,
floatfix,
longbibliography
]{revtex4-2}

\usepackage{times}
\usepackage{amstext}
\usepackage{graphicx}
\usepackage{color}
\usepackage{soul}
\usepackage{MnSymbol}
\usepackage{subfiles}
\usepackage{bm}
\usepackage{hyperref}
\usepackage{layouts}
\usepackage{enumitem}
\usepackage{natbib}
\usepackage{comment}
\usepackage{xcolor}

\begin{document}

\title{From Random Walks to Thermal Rides: Universal Anomalous Transport  in Soaring Flights}

\author{Jérémie Vilpellet}

\affiliation{EconophysiX Lab, Institut Louis Bachelier, 28 Pl.~de la Bourse, Palais Brongniart, 75002 Paris, France}
\affiliation{LadHyX UMR CNRS 7646, École Polytechnique, Institut Polytechnique de Paris, 91128 Palaiseau Cedex, France}

\author{Alexandre Darmon}
\affiliation{Art in Research, 33 rue Censier, 75005 Paris, France}

\author{Michael Benzaquen}\email{michael.benzaquen@polytechnique.edu}
\affiliation{EconophysiX Lab, Institut Louis Bachelier, 28 Pl.~de la Bourse, Palais Brongniart, 75002 Paris, France}
\affiliation{LadHyX UMR CNRS 7646, École Polytechnique, Institut Polytechnique de Paris, 91128 Palaiseau Cedex, France}
\affiliation{Capital Fund Management, 23 Rue de l’Université, 75007 Paris, France}

\begin{abstract}

Cross-country soaring flights rely on intermittent atmospheric updrafts to cover long distances, producing trajectories that alternate between rapid relocation and local exploration.
From a large dataset of paraglider, hang glider, and sailplane flights, we uncover a universal transport law: beyond short ballistic times, horizontal motion is persistently sub-ballistic, with a Hurst exponent 
$\approx 0.88$ largely independent of aircraft type. Phase-resolved analysis using a probabilistic segmentation method shows that this scaling arises from the fundamentally intermittent, two-dimensional, and directionally correlated nature of soaring transport, in which successive ballistic segments do not add coherently. We find that learning, in the sense of experience-driven improvements in exploration and decision-making, manifests primarily in the search phase, enhancing the ability to efficiently probe the air mass and locate the next thermal.
Overall, our results suggest that atmospheric structure and the generic organization of the transition–search–climb cycle dominate transport properties, placing human soaring alongside biological and physical systems where anomalous transport emerges from intermittency and persistence.
    
\end{abstract}

\date{\today}

\maketitle

\section{Introduction}

Long-range movement, from migrating animals to foraging predators and traveling humans, unfolds across scales and disciplines in environments that are structured yet uncertain, and patchy in both space and time \cite{turchin1998quantitative,codling2008random,Nathan2008MovementEcologyParadigm}. A particularly striking example is soaring, where the “resource” is atmospheric energy itself~\cite{reichmann1993cross}.
Birds and soaring aircraft alike exploit columns of rising air by circling to gain altitude and then gliding onward until lift is encountered again~\cite{pennycuick2008modelling,reichmann1993cross}. 
Such patches of lift, known as \textit{thermals}, are buoyant updrafts that form when sunlight heats the ground. The warmed surface then heats the air in contact with it, triggering rising plumes that collectively constitute the daytime convective boundary layer. These plumes are spatially and temporally intermittent, rising until they lose buoyancy through thermalization with the surrounding air and typically dispersing near the boundary-layer top or {cloud base}. In practice, thermals are advected and distorted by the ambient flow and are therefore typically tilted rather than forming perfectly vertical columns~\cite{stull2012introduction}.

Over the past decades, it has become increasingly clear that animal movement in natural environments is far from trivial~\cite{Nathan2022BigDataAnimalMovement,Kareiva1983CorrelatedRandomWalk,Mueller2013SocialLearningMigratory}. Early tracking analyses reported Lévy-walk-like statistics in wandering albatross flights, a result that helped catalyze a broad literature on scale-free movement and search~\cite{Viswanathan1996LevyAlbatross,Reynolds2015LiberatingLevy}. Subsequent work revisited—and in some cases challenged—these conclusions using higher resolution data and likelihood-based model comparison, highlighting how inferred “laws” of movement can depend sensitively on sampling resolution and statistical methodology~\cite{Edwards2007RevisitingLevy,clauset2009power,Reynolds2018CurrentStatusLevy}. Along similar lines, large multi-species datasets of marine predators, including sharks and tunas, have revealed switches between Lévy-like and Brownian modes depending on environmental context~\cite{Sims2008ScalingLaws,Humphries2010EnvironmentalContext,zaburdaev2015levy}. More recently, ultra-high-resolution tracking has uncovered further departures from simple  descriptions, including signatures of ergodicity breaking in the area-restricted search behavior of avian predators~\cite{vilk2022ergodicity}. 
A key enabler of these advances is scale: modern GPS and biologging technologies now generate population-level trajectory archives, enabling robust statistical comparisons across individuals, environments, and even experience levels~\cite{Kays2015TerrestrialTracking,Kays2022MovebankSystem,Nathan2022BigDataAnimalMovement}. For example, GPS-based analyses of soaring raptors show that adults outperform juveniles under challenging thermal-soaring conditions, consistent with skill-dependent improvements in thermal centering~\cite{harel2016adult,Ruaux2020FlightBehaviours}. {Along the same lines, work on terrestrial migrants such as ungulates has revealed strongly directed seasonal routes shaped by resource phenology (green-wave surfing) and by learned, socially transmitted migratory knowledge~\cite{Jesmer2018UngulateCulture,Aikens2017Greenscape}}. 

Comparable trajectory archives are now available for human soaring. Affordable GPS devices and widespread flight logging have produced large repositories of paraglider, hang gliders and sailplane tracks~\cite{cfdffvl,netcoupe,xcontest}, thereby enabling similarly rich statistical investigations of human movement in an atmospheric “resource landscape”.
That said, a robust, large-scale quantitative characterization of transport in real cross-country soaring is, to our knowledge, still lacking. In particular, it is unclear which mechanisms ultimately govern large-scale transport, and how transport properties arise from the combined effects of aircraft dynamics, pilot decision-making, and atmospheric variability. 
While soaring trajectories clearly contain ballistic-like segments during glides, it is unclear what global transport law one would expect once the full intermittency of the flight cycle is taken into account~\cite{metzler2000random,benichou2011intermittent}. We therefore ask: What are the emergent large-scale transport laws of cross-country soaring? Do they differ across aircraft types? And can large trajectory datasets reveal systematic signatures of experience, namely learning effects, in how pilots search for and exploit atmospheric lift?

Here, we address these questions using a large dataset of publicly available GPS trajectories spanning paragliders, hang gliders, and sailplanes. We first characterize global transport scaling using standard dispersion diagnostics from transport physics and movement ecology, enabling direct comparisons across aircraft categories. We then introduce a phase-resolved analysis by segmenting flights into transition, search, and climb phases using a Hidden Markov Model built from simple trajectory-derived features, allowing us to relate global transport properties to the intermittent structure of soaring flight. Focusing on paragliding, where proxies for pilot skill are most readily available, we quantify how experience modulates flight statistics, associating “learning” primarily to how efficiently lift is detected and exploited during the search phase. Finally, we interpret our results through the lens of intermittency, directional persistence, and correlated random-walk theories, and outline directions for future research.

\section{Data}

Our dataset {consists} of publicly available soaring aircraft trajectories recorded in France between 2016 and 2021. In particular we have access to {78,645} trajectories for paragliders, 2,656 for flex-wing hang gliders, and 21,295 for sailplanes from~\cite{cfdffvl} and~\cite{netcoupe}. The trajectories are provided in natural geodetic coordinates $(\phi_t,\lambda_t,z_t)$, where $\phi_t$ denotes latitude, $\lambda_t$ longitude, and $z_t$ the GPS-derived ellipsoidal height (WGS–84) or \textit{GPS altitude}~\cite{hofmann2008gnss,torge2023geodesy}. To perform statistical analyses in a locally Euclidean frame, we convert these  coordinates into local ENU (East–North–Up) Cartesian coordinates $\boldsymbol r_t =(x_t,y_t,z_t)$~\cite{farrell1999gps}. The ENU system is defined as the tangent-plane coordinate system centered at a fixed reference point, with the $x$-axis pointing East and the $y$-axis pointing North. In this frame, the $(x,y)$-plane corresponds to the local East–North horizontal plane, and $z$ retains the physical interpretation of altitude.

Among all trajectories, we retain flights sampled at a minimum frequency of $1\,\mathrm{Hz}$, lasting at least one hour, free of spatial or temporal discontinuities, and spanning a horizontal range of at least $15\,\mathrm{km}$—that is, roughly an order of magnitude larger than the typical altitude variations observed during a regular cross-country soaring flight. After this filtering step, 49,939 paraglider trajectories, 780 hang-glider trajectories, and 9,203 sailplane trajectories remain, and all statistical analyses presented below are based on this filtered dataset.

\section{Global transport scaling}

To quantify transport dynamics in soaring flights, we rely on the mean square displacement (MSD) computed in the East–North horizontal plane, a classical diagnostic of dispersal and spreading in physical stochastic processes~\cite{metzler2000random} and movement ecology~\cite{turchin1998quantitative,codling2008random}. The MSD measures how far, on average, the aircraft has moved over a time interval $\Delta$, and is defined by:
\begin{equation}
    \delta^2(\Delta) = \mathbb E\left[ (\boldsymbol{x}_{t+\Delta} - \boldsymbol{x}_t)^2 \right],\label{eq:MSD}
\end{equation}
where $\boldsymbol x_t= (x_t,y_t)$, and where we have assumed that the underlying process is stationary and second-order (finite variance). 

\begin{figure}[t!]
    \centering
\includegraphics[width=\linewidth]{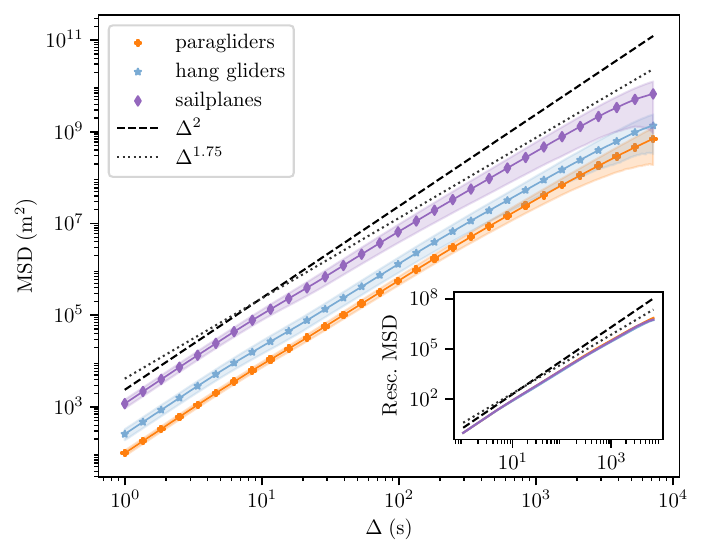}
    \caption{Mean Square Displacement (MSD) as function of time lag $\Delta$, as defined in Eq.~\eqref{eq:MSD},  for paragliders, hang gliders and sailplanes.  The  shaded areas indicate standard deviations computed over all trajectories. The inset shows the MSD rescaled by the average instantaneous velocity for each glider type, highlighting their near-perfect collapse and, consequently, the universal nature of their global transport properties.}
    \label{fig:vario_global}
\end{figure}

Figure~\ref{fig:vario_global} displays the MSD for paragliders, hang gliders, and sailplanes. The three curves differ only by a vertical shift reflecting their distinct characteristic speeds. This indicates that paragliders, hang gliders, and sailplanes alike exhibit essentially the same underlying transport dynamics. On very short timescales ($\Delta < 10\,\mathrm{s}$), one has $\delta^2 \sim \Delta^{2}$, consistent with ballistic motion, which can be attributed to the limited maneuverability of gliders over such short time intervals. On larger timescales ($\Delta > 10\,\mathrm{s}$), the system transitions  to a sub-ballistic, superdiffusive regime $\delta^2 \sim \Delta^{2H}$ with $\frac12<H\approx 0.88 < 1$, where $H$ denotes the Hurst exponent \footnote{For fractional Brownian motion, and more generally for $H$-self-similar processes, the Hurst exponent $H\leq1$ is defined by the scaling $\delta^2\sim\Delta^{2H}$. Values $2H>1$ correspond to superdiffusive (persistent) behavior, $2H<1$ to subdiffusive (antipersistent) behavior, and $2H=1$ to normal diffusion (Brownian motion). In practice, the MSD scaling may be time-dependent, with different effective exponents observed at different timescales.}. 
In the following section, we attempt to shed light on the origin of such universal superdiffusive transport dynamics in cross-country soaring flights.

\section{Transition, search and climb}

Glider aircrafts during cross-country soaring flights typically progress through a repeating sequence of two or three flight phases~\cite{allen2005autonomous,pennycuick2008modelling}. When leaving a thermal, they  glide toward the next source of lift, a \textit{transition} phase during which they follow a relatively straight, energy-efficient trajectory governed primarily by aerodynamics and gravity. This continues until the pilot detects the presence of rising air~\footnote{Note that experienced pilots, when at sufficiently high altitude, may deliberately bypass thermals encountered during a glide in order to minimize total flight time, anticipating more favorable lift later along their route.}. But if no thermal is encountered at reasonable altitude above ground level, the glider enters a \textit{search} phase, in which the objective is to explore a sufficiently large area to locate the next usable thermal while minimizing altitude loss. During this phase, the pilot often performs gentle turns, course adjustments, or probing maneuvers aimed at sampling the surrounding air mass without descending too rapidly to avoid an unintended landing before a new lift source is found. Once a thermal is located, the glider enters into the \textit{climb} phase, circling within the rising air to regain altitude.  After the climb is completed, the cycle repeats. 
\begin{figure}[t!]
    \centering
    \includegraphics[width=\linewidth]{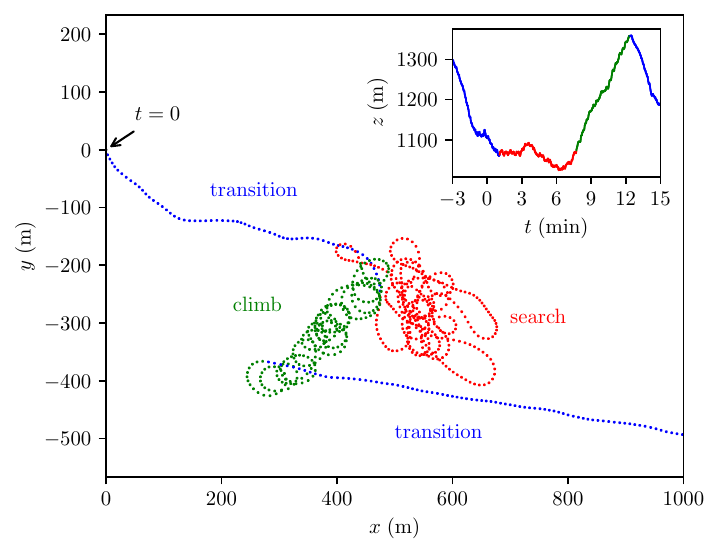}
    \caption{Example of paraglider trajectory in the $xy$-plane, illustrating the different phases of the soaring  cycle, as obtained from the HMM estimation procedure: transition (blue), search (red), and climb (green). The inset displays the altitude as function of time.}
    \label{fig:para_trajectory}
\end{figure}

Figure~\ref{fig:para_trajectory} displays a segment of a paragliding flight in which the transition, search, and climb phases can be clearly identified. This trajectory reveals that each phase exhibits distinct transport properties, and that their succession may collectively account for the global transport dynamics discussed in the previous section.

Developing a systematic and robust method to distinguish between these flight phases is therefore essential for a deeper understanding of soaring behavior. The challenge stems from the fact that these phases are only partially objective. While the example shown in Figure~\ref{fig:para_trajectory} offers a textbook case where the three phases can be clearly distinguished, most trajectories are far less straightforward to classify. In many flight segments, the boundaries between phases are blurred, and it is often ambiguous whether a given portion of the trajectory should be classified as a transition or a search phase, or even as the onset of a climb.
The most realistic goal is therefore to develop a segmentation method that produces convincing results across a wide variety of situations and is thus likely to generalize well.
To achieve this, we adopt a probabilistic framework based on the following idea: each glider trajectory can be viewed as a realization of an observable discrete-time stochastic process driven by an unobserved (latent) state variable, where each state corresponds to a specific flight phase. Within each phase, pilots tend to exhibit characteristic behaviors that imprint distinguishable features on the observed trajectory, naturally motivating the use of a state-space model. By further assuming that the latent state evolves as a Markov chain and that the observations are conditionally independent and stationary given the state, we obtain a {H}idden Markov {M}odel (HMM)~\cite{rabiner2002tutorial}, which can be readily calibrated on our dataset to discriminate between the different flight phases.
However, raw trajectory observations at each time step carry limited information about the underlying phase, and they clearly fail to satisfy conditional independence and stationarity. We therefore transform the trajectories into a set of more informative features that better adhere to the HMM assumptions. These include binary variables such as the sign of the vertical speed, curvature-angle persistence, and indicators of trajectory straightness. The full estimation procedure is detailed in the appendix.

Figure~\ref{fig:vario_regime} displays the  conditional MSD for each flight phase. As expected, the transition phase exhibits ballistic behavior at all timescales ($H=1$) for the three aircraft types. The search and climb phases also show ballistic behavior at short timescales ($\Delta < 10\,\mathrm{s}$), but reveal more intricate dynamics at larger scales ($\Delta > 10\,\mathrm{s}$). The search phase, in particular, displays strongly subdiffusive behavior, with $H \approx 0.3$ for paragliders and hang gliders, and $H \approx 0.2$ for sailplanes. This phase—during which pilots steer through gentle turns to sample the air mass and pinpoint the next thermal—thus contributes significantly to ``slowing down" the global transport dynamics. 
The climb phase exhibits quasi-ballistic scaling at long timescales, but with a very small effective velocity (almost two orders of magnitude lower than the characteristic speeds of the aircrafts). This reflects the fact that thermal updrafts are almost never vertical but typically tilted by the wind or the breeze~\cite{stull2012introduction,reichmann1993cross}, causing pilots to drift slowly while climbing. In the case of sailplanes, the radii of curvature are so large (see below) that the crossover from short-time ballistic behavior to its longer-timescale counterpart shows a transient negative slope, indicating a brief regime of negative directional correlation at intermediate timescales (10–20\,s). Such timescales are consistent with the typical turning period: 
a full revolution during thermalling takes on average slightly less than 30\,s, so that anticorrelations naturally emerge as the aircraft begins to curve back on itself. Note that such typical turning period is shared by all three glider aircraft types.

\begin{figure*}[t!]
    \centering
    \includegraphics[width=\linewidth]{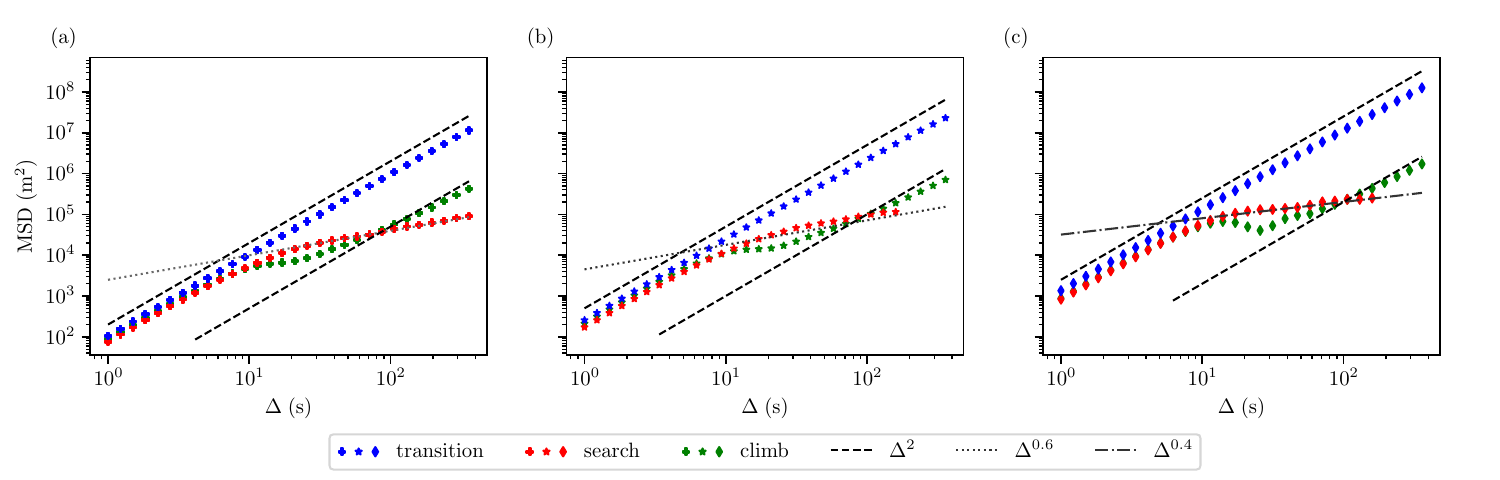}
    \caption{Conditional Mean Square Displacement (MSD) as function of time lag $\Delta$, per flight phase, for (a) paragliders, (b) hang gliders, and (c) sailplanes. The color code is the same as in Fig.~\ref{fig:para_trajectory}, also used consistently in Figs.~\ref{fig:vario_global_regime_para} and \ref{fig:prop_regime_flight}: transition (blue), search (red), and climb (green).}
    \label{fig:vario_regime}
\end{figure*}

\begin{figure*}[t!]
    \centering
    \includegraphics[width=\linewidth]{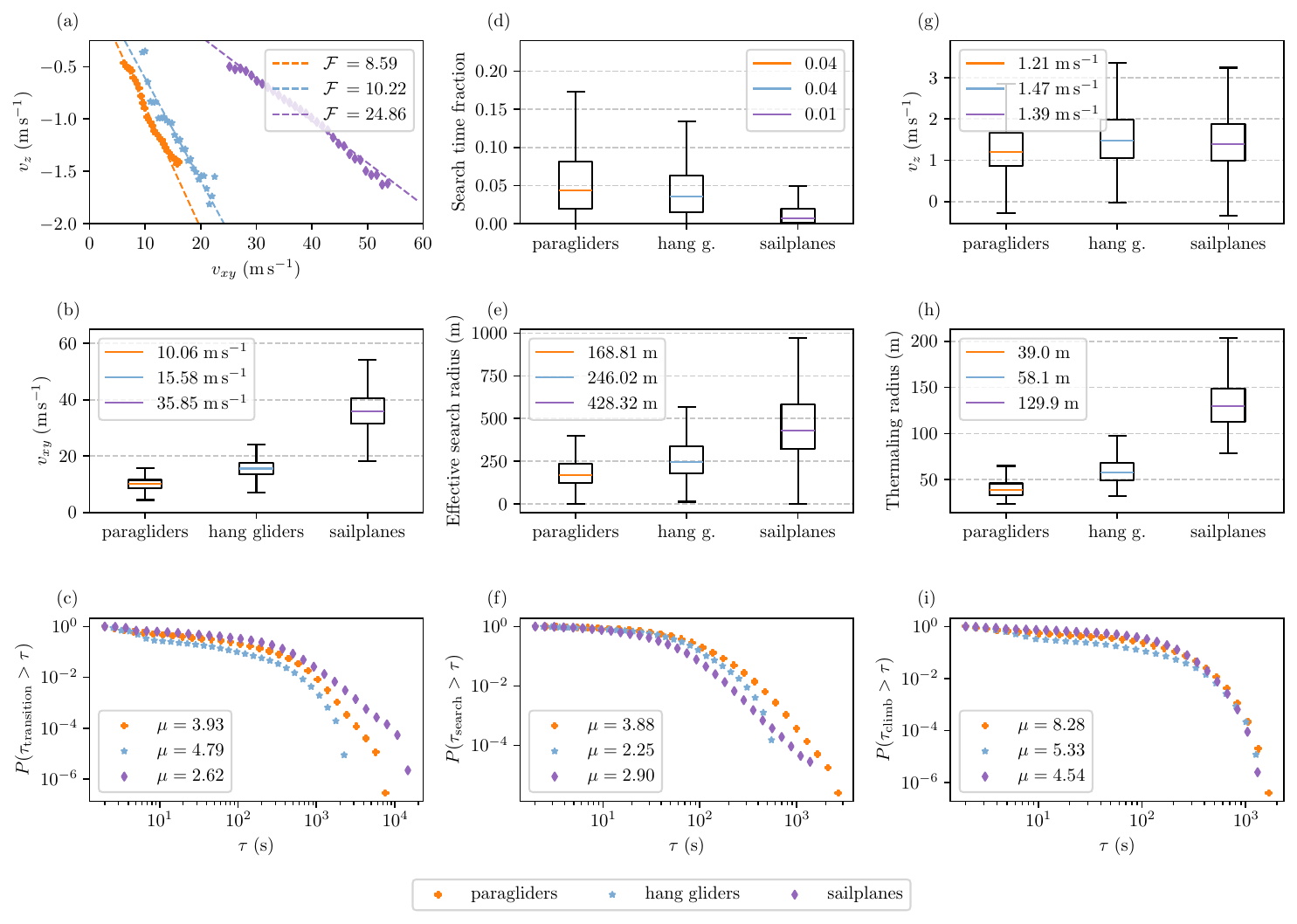}
    \caption{Transport statistics across glider types, for the different phases: transition (left column), search (center column), and climb (right column). (a) Sink rate as function of horizontal velocity, where the fit $v_{xy}=\mathcal F  |v_z|$ defines the glide ratio $\mathcal F$. (b) Box plots of horizontal velocities. (c) Survival function of transition times. (d) Box plots of search time fractions. (e) Box plots of effective search radii. The effective search radius is computed as the radius of the smallest circle in the $xy$-plane  containing the search trajectory. (f) Survival function of search times. (g) Box plot of climb rates. (h) Box plots of thermaling radii. (i) Survival function of thermaling times. Tail exponents $\mu$, defined through $ P(\bullet >\tau) \sim \frac{1}{\tau^{\mu}}$, are computed using the Hill estimator, see~\cite{clauset2009power}. Box plots show the median (central line), the interquartile range or IQR (box, spanning the 25th to 75th percentiles), and whiskers extending to the most extreme data points within $1.5\times$IQR from the quartiles. } 
    \label{fig:statistics_gliders}
\end{figure*}

Figure~\ref{fig:statistics_gliders} presents descriptive statistics that highlight the distinctive features of each phase. The transition phase is  well suited for estimating real glide ratios~\cite{anderson2005introduction,reichmann1993cross} (Fig.~\ref{fig:statistics_gliders}a). We obtain ${\mathcal{F}\approx  8.5}$ for paragliders, $\mathcal{F}\approx 10$ for hang gliders, and $\mathcal{F}\approx 25$ for sailplanes. The statistics of  horizontal velocities are displayed in panel~\ref{fig:statistics_gliders}b. The distribution of transition times (Fig.~\ref{fig:statistics_gliders}c) is  noticeably more heavy-tailed for sailplanes than for paragliders and hang gliders, which can most likely be attributed to their much higher glide ratios allowing for some extreme transitions.

\begin{figure}[h!]
    \centering
    \includegraphics[width=\linewidth]{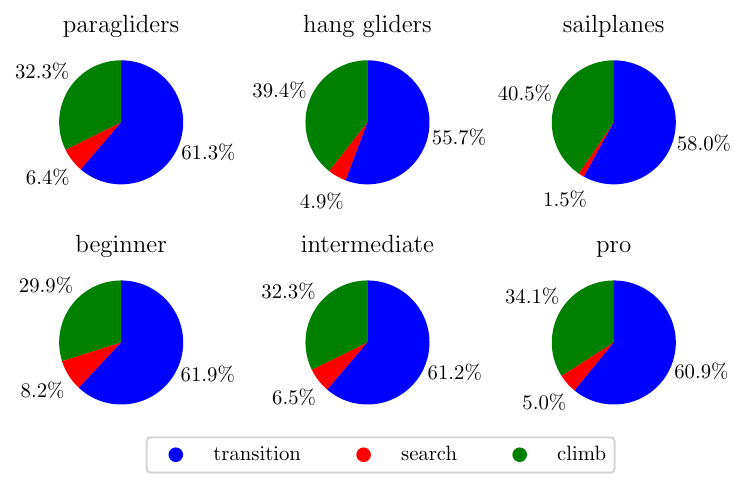}
    \caption{Pie charts showing the average fractions of flight time devoted to the different phases. Top row: different glider aircraft types. Bottom row: paraglider categories. 
    }
    \label{fig:prop_regime_flight}
\end{figure}

The time spent in the search phase (Fig.~\ref{fig:statistics_gliders}f) exhibits the thickest tails for all gliders, consistent with the wide variety of search conditions encountered. These include terrain nature and topography, cloud indicators, as well as dynamic lift generated by wind interacting with the relief (orographic lift), which can locally sustain flight, modify search strategies, and significantly prolong or shorten the search phase. The fraction of flight time devoted to searching (Figs.~\ref{fig:statistics_gliders}d and \ref{fig:prop_regime_flight}) is, on average, highest for paragliders and hang gliders, whereas sailplanes appear markedly more efficient, with only about 1-2\% of the flight spent in this phase. This increased efficiency can again be attributed to their much larger glide ratio and horizontal velocity, allowing them to cover substantially larger areas much more rapidly than paragliders and hang gliders (see effective search radii in Fig.~\ref{fig:statistics_gliders}e). The fraction of time spent in the other phases shows much less variation between the different glider types, with $\approx 60\%$ of the time spent transitioning, and $\approx 35\%$ spent climbing (see Fig.~\ref{fig:prop_regime_flight}).

Vertical climb velocities (Fig.~\ref{fig:statistics_gliders}g) are quite similar across glider types, consistent with their comparable sink rates. Differences become apparent, however, when examining thermaling radii (Fig.~\ref{fig:statistics_gliders}h), which reflect their distinct maneuvering capabilities and determine the range of thermal structures that can be effectively exploited. Paragliders, for instance, can tighten their turns sufficiently to make use of the narrowest thermal updrafts. Finally, the distribution of climbing times (Fig.~\ref{fig:statistics_gliders}i) is very thin-tailed and is best fitted by an exponential, consistent with the fact that climbs are ultimately limited by the cloud-base altitude~\cite{wallace2006atmospheric}.

 \section{Learning}

To investigate learning we focus on paragliders, for which we have the largest dataset and for which correlating pilot skill level with glider class is most straightforward. We group our data into three bins according to the EN (European Norm) certification~\cite{EN926}, which classifies paraglider wings from A (highest passive safety, beginner level) to D and CCC~\cite{FAI_CCC} (competition wings with demanding handling). We define three broad categories: “beginner” for EN–A and EN–B\textsuperscript{–} wings (12\% of all trajectories), “sports class” for EN–B\textsuperscript{+} and EN–C wings (61\%), and “performance/competition” for EN–D and CCC wings (27\%).
Although this classification may not be perfectly aligned with pilot skill—since some pilots may occasionally fly wings above or below their true level—it remains a very reasonable proxy for pilot experience and flying style at the population scale.

\begin{figure}[t!]
    \centering
    \includegraphics[width=\linewidth]{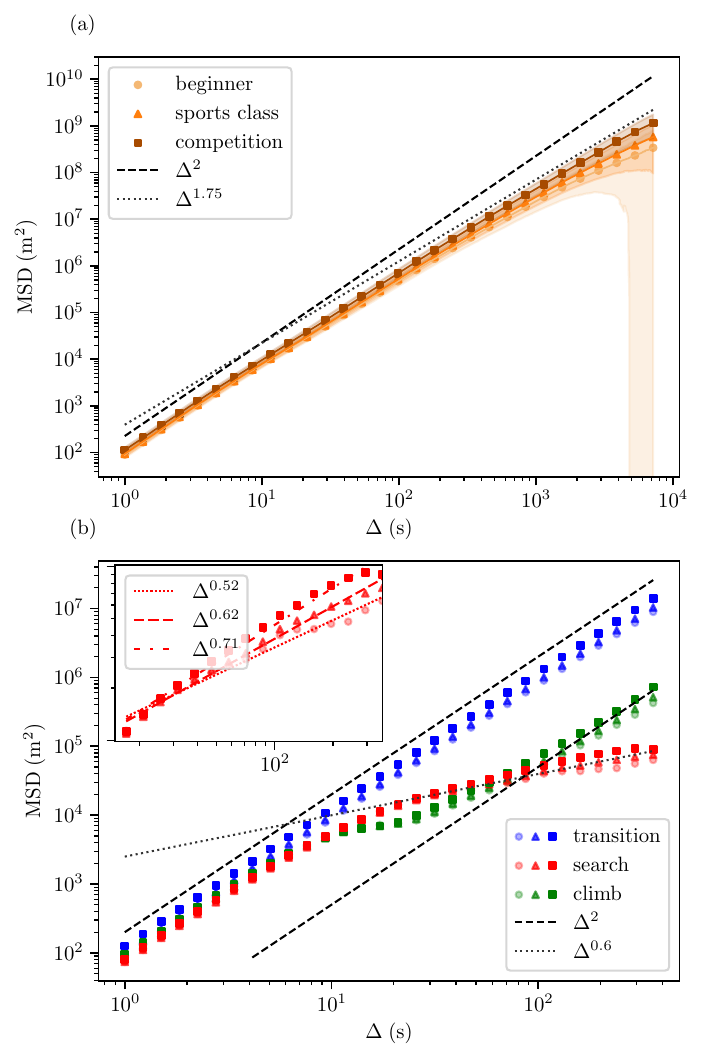}
    \caption{Paragliders Mean Square Displacement (MSD) as function of time lag $\Delta$ for beginner ($\bullet$), sports class ($\scriptstyle \blacktriangle$) and competition wings (${\scriptstyle{\blacksquare}}$). (a) Global. (b) Per flight phase. The inset zooms into the long-time regime of the search phase to stress asymptotic scaling differences.}
    \label{fig:vario_global_regime_para}
\end{figure}

Figure~\ref{fig:vario_global_regime_para}a displays the MSD for these three categories separately and reveals clear evidence of learning: the competition class exhibits a steeper slope than the sports class, which in turn is steeper than the beginner category. Since the slope of the MSD directly reflects the persistence and efficiency of horizontal transport, a steeper slope indicates more effective exploitation of atmospheric structures and, therefore, more proficient cross-country performance. 
More specifically, the global Hurst exponent shows a statistically significant increase across categories, rising from $H \approx 0.85$ for beginners to $H \approx 0.92$ for competition gliders.  

\begin{figure*}[t!]
    \centering
    \includegraphics[width=\linewidth]{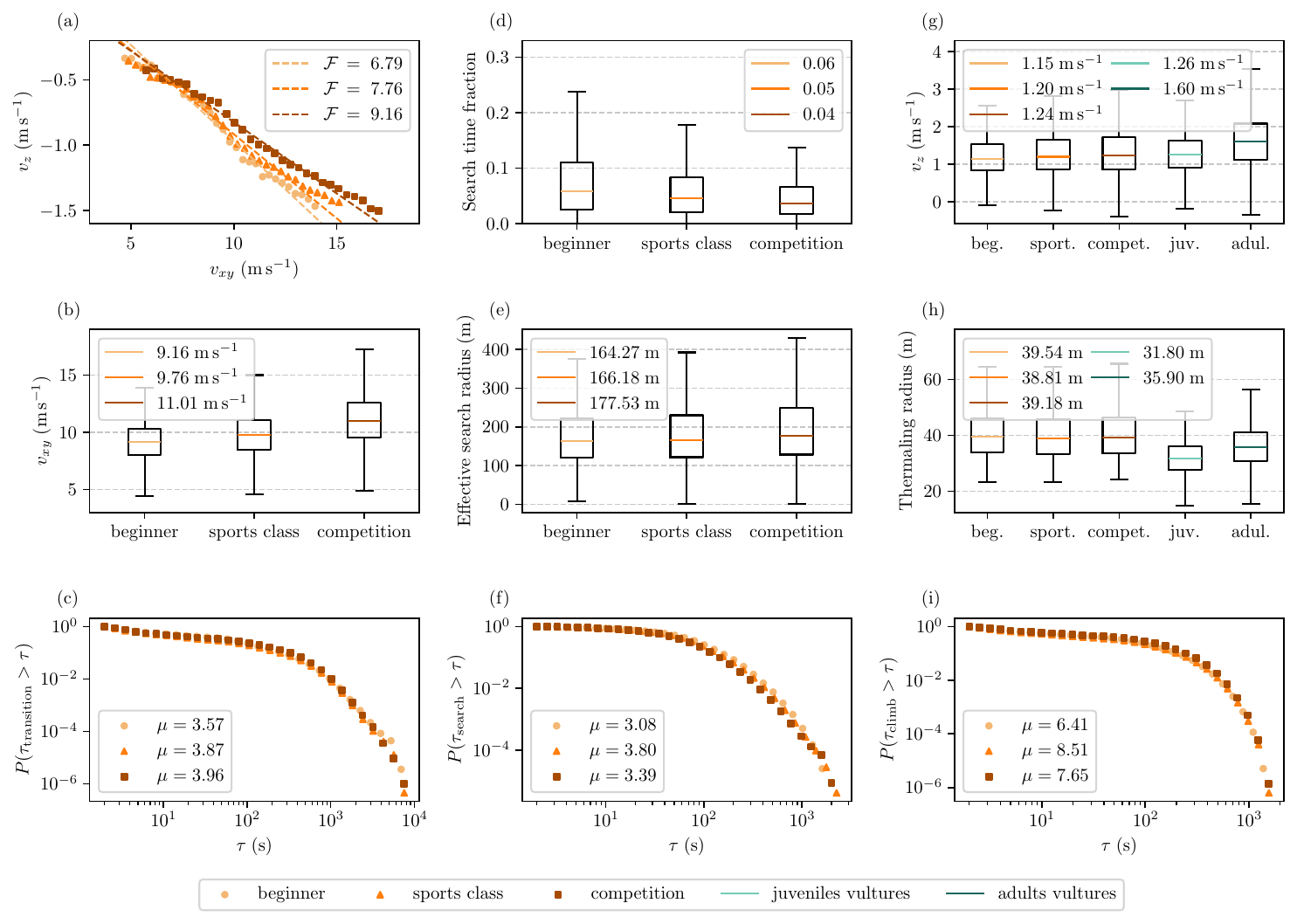}
    \caption{Paragliders transport statistics across wing classes, for the different phases: transition (left), search (center), and climb (right). (a) Sink rate as function of horizontal velocity, where the fit $v_{xy}=\mathcal F  |v_z|$ defines the glide ratio $\mathcal F$. (b) Box plots of horizontal velocities. (c) Survival function of transition times. (d) Box plots of search time fractions. (e) Box plots of effective search radii. (f) Survival function of search times. (g) Box plot of climb rates. (h) Box plots of thermaling radii. (i) Survival function of thermaling times.
    In panels (g) and (h) we have include values for juvenile and adult vultures from~\cite{harel2016adult} for comparison.
    }
    \label{fig:statistics_learning}
\end{figure*}

In order to understand where learning plays the most critical role, we plot the MSD separately for each flight phase and for each of the three glider categories (Fig.~\ref{fig:vario_global_regime_para}b). The vertical shifts observed in the transition and climb phases primarily reflect differences in wing performance rather than learning effects: glide ratios increase markedly with glider class (Fig.~\ref{fig:statistics_learning}a), as do average horizontal velocities (Fig.~\ref{fig:statistics_learning}b),
the latter also benefiting from the more efficient and extended use of the speed system on higher-performance wings.
In contrast, the search phase exhibits statistically significant differences across categories, with Hurst exponents increasing systematically with glider class, from $H \approx 0.26$ for beginners, to $H \approx 0.31$ for the sports class, and up to $H \approx 0.35$ for competition gliders. This result indicates that learning manifests most strongly during the search phase: In other words, experience and skill primarily enhance the ability to actively probe, interpret, and exploit atmospheric cues, rather than merely improving ballistic flight or climbing performance (although the latter does improve moderately with glider class, see Fig.~\ref{fig:statistics_learning}g). This increased search efficiency is further reflected in the fraction of flight time devoted to searching, which decreases systematically with glider class (Figs.~\ref{fig:statistics_learning}d and \ref{fig:prop_regime_flight}). 
The effective search radius also increases slightly with glider class (Fig.~\ref{fig:statistics_learning}e), consistent with more efficient spatial exploration during the search phase.
While, as noted above, climb velocity improves with glider class, we find no consistent correlation between thermaling radius and pilot skill level. This result contrasts with observations in vultures, for which a systematic increase in vertical climb velocity is observed between juvenile and adult individuals (Fig.~\ref{fig:statistics_learning}g). Interestingly, both juvenile and adult vultures appear more efficient climbers than paragliders. In vultures, however, this improvement is associated with differences in thermaling radius (Fig.~\ref{fig:statistics_learning}h), which is smaller for both age classes than for paragliders—a consequence of their smaller size and superior {maneuvrability}.
According to the authors of Ref.~\cite{harel2016adult}, adult birds climb more strongly but circle with wider radii than juveniles. Younger birds fail to widen their turns to compensate for the degradation in glide efficiency caused by increased bank angle. This ability is gradually acquired through learning.
Finally, we find no significant differences in the tails of the distributions of transitioning, searching and climbing times (Figs.~\ref{fig:statistics_learning}c, \ref{fig:statistics_learning}f and \ref{fig:statistics_learning}i).

In summary, increased pilot skill and experience effectively raise the thermal detection rate (or detection range) during search, and improve the trade-off between exploration and altitude loss, thereby reducing the time spent searching and increasing net transport persistence.

\section{Discussion}

Let us now return to our initial empirical finding: robust, universal sub-ballistic, superdiffusive horizontal transport with a Hurst exponent $H \approx 0.88$ (Fig.~\ref{fig:vario_global}), observed consistently across paragliders, hang gliders, and sailplanes despite large differences in speed, glide ratio, and maneuverability.

The phase-resolved analysis presented above shows that flights can be decomposed into a repeating sequence of transition, search, and climb segments. Transition segments are individually ballistic, search phases exhibit subdiffusive behavior, and climb segments can, to a good approximation, be treated as non-contributing to horizontal transport: although technically ballistic, their drift velocities are much smaller than characteristic horizontal speeds, effectively rendering them immobile phases of finite duration.

A naive one-dimensional argument would then suggest that ballistic segments should dominate the long-time behavior, leading asymptotically to $\delta^2(\Delta)\sim\Delta^2$. Indeed, if the ballistic direction were persistent, one could decompose the displacement over a window of size $\Delta$ as
 $x_{t+\Delta}-x_t = v\,T_{\mathrm b}(\Delta) + x_{\mathrm s}(\Delta)$, 
where $v$ is the velocity during ballistic phases, $T_{\mathrm b}(\Delta)$ the total time spent in ballistic motion within the window, and $x_{\mathrm s}$ the cumulative displacement during subdiffusive phases (immobile phases do not contribute). Assuming that the subdiffusive phases have zero mean and that their increments are independent of the ballistic durations yields 
 $\delta^2(\Delta)=v^2\,\mathbb E[T_{\mathrm b}^2] + \mathbb E[x_{\mathrm s}^2]$. 
Denoting by $\eta_{\mathrm b}$ the ballistic fraction of the flight time, one obtains $\delta^2\sim(\eta_{\mathrm b}v)^2\Delta^2$, independently of the immobile or subdiffusive phases, which only affect subleading terms. This prediction is clearly incompatible with the observed value $H\approx0.88$, highlighting the limitations of this simple picture and calling for a more nuanced interpretation.

A key implicit assumption underlying the one-dimensional argument is that ballistic displacements add coherently along a fixed direction, producing a net drift that grows linearly in time and whose square therefore dominates the MSD at long times. In real soaring flights, however, this assumption is violated because the transport dynamics are intrinsically two-dimensional. Even during transition phases, the heading of the aircraft is continually adjusted in response to terrain, meteorological cues, airspace constraints, and strategic decisions made by the pilot. Search and climb phases further introduce strong and often rapid reorientations. As a result, successive ballistic segments are generally not collinear, and their vectorial contributions to the net displacement partially cancel.

More formally, the MSD in two dimensions can be expressed using the Green--Kubo relation~\cite{kubo1985statistical},
 $\delta^2(\Delta)=2\int_0^{\Delta}(\Delta-t)\,C_{\boldsymbol v}(t)\,dt$, 
where $C_{\boldsymbol v}(t)=\mathbb E[\boldsymbol v_{t'}\cdot\boldsymbol v_{t'+t}]$ is the horizontal velocity autocorrelation function. Pure ballistic motion corresponds to a constant $C_{\boldsymbol v}(t)$, yielding $\delta^2\sim\Delta^2$. Any decay of velocity correlations---due to reorientation, looping, or strategy changes---necessarily reduces the exponent. The observed value $H\approx0.88$ thus directly signals long-lived but imperfect persistence in the horizontal velocity.

To gain further insight, we now examine possible mechanisms that can naturally account for the observed sub-ballistic superdiffusion. A first candidate is L\'evy-walk--type transport, emerging from the presence of heavy-tailed transition times~\cite{zaburdaev2015levy}. As discussed above, soaring flights are inherently intermittent: long transition phases are interspersed with search and climb phases that contribute little to net horizontal displacement. If the durations of transition phases are heavy-tailed, as suggested by Fig.~\ref{fig:statistics_gliders}c for sailplanes ($\mu<3$, infinite variance), and if the direction of motion is at least partially randomized between transitions, the resulting dynamics resemble a L\'evy walk with finite velocity~\cite{zaburdaev2015levy,metzler2000random}. In this regime, one expects $\delta^2\sim\Delta^{4-\mu}$, consistent with $H<1$. The observed value $H\approx0.88$ would correspond to $\mu\approx2.25$, well within the range reported for intermittent transport processes in physics and biology~\cite{codling2008random}.

To this mechanism may be added the effect of finite orientational persistence within ballistic phases. As suggested above, in practice, transition phases are rarely perfectly straight: the aircraft speed may remain approximately constant while the heading undergoes slow rotational diffusion. This situation is well described by persistent random walks or active Brownian motion models, in which motion is ballistic at short times and diffusive at long times, with a broad crossover regime exhibiting an apparent superdiffusive exponent~\cite{howse2007self,bechinger2016active}. If the persistence time is comparable to the duration of transition phases, this effect alone can reduce the effective MSD exponent below~2 and naturally leads to extended crossover regimes.

More generally, these considerations point to pre-asymptotic scaling~\cite{meroz2015toolbox}. Even if a small directional bias exists (for instance due to wind or route optimization), the MSD may contain competing contributions,
 $\delta^2(\Delta)\sim a\,\Delta^{2}+b\,\Delta^{\gamma}$, where $\gamma<2$, 
with the ballistic term dominating only at timescales larger than those accessible in finite trajectories. Over an intermediate range of lags, a single power-law fit can therefore yield an effective exponent $H<1$~\cite{metzler2000random,meroz2015toolbox}. In other words, a mixture of contributions from phases with different scaling properties naturally produces an effective power law with an exponent below 2, even if a truly ballistic regime might exist asymptotically beyond observable timescales in real flights.

Finally, averaging effects may further contribute to sub-ballistic scaling. MSDs are typically estimated from time averages along single trajectories, and in intermittent processes with broad sojourn times, time-averaged and ensemble-averaged MSDs need not coincide, even in the long-time limit~\cite{he2008random,lubelski2008nonergodicity}. Such weak ergodicity breaking can bias exponent estimates and should therefore be kept in mind when interpreting finite-dataset scaling laws. See also~\cite{vilk2022ergodicity,vilk2025strong} for applications in a somewhat related topic.

\section{Concluding remarks}

Taken together, the considerations presented in the previous section show that the observed exponent $H\approx 0.88$ does not contradict the existence of ballistic flight phases. Rather, it reflects the fundamentally {intermittent, two-dimensional, and directionally correlated} nature of soaring transport. The near universality of the exponent across glider types further suggests that large-scale atmospheric structure and the generic organization of the transition--search--climb cycle play a more decisive role than aircraft-specific performance.

Several directions naturally follow from this work. First, direct measurements of heading and velocity autocorrelations, both globally and conditioned on flight phase, would allow one to discriminate between Lévy-walk-like dynamics and finite-persistence correlated random walks. Second, a step-based description in which displacements between successive climbs are treated as elementary transport events could provide a natural framework for testing Lévy-walk predictions and connecting scaling exponents to the statistics of thermal spacing. Third, separating air-relative from ground-relative motion would help disentangle pilot strategy from atmospheric advection. Finally, extending this analysis to different meteorological regimes or geographical regions may clarify how environmental heterogeneity shapes large-scale transport properties.
{A further line of investigation would be to examine how transport properties differ between solo and collective flights. 
In paragliding competitions, pilots effectively engage in time-based races along predefined tasks, with staggered start gates and strong interactions mediated by visual and tactical cues. Starting early requires relying solely on one’s own interpretation of terrain and aerology, whereas starting later allows pilots to exploit social information, using the trajectories of others to infer the structure of the air mass (lift and sink areas, optimal lines, etc.)~\cite{reichmann1993cross}. However, starting too late may limit the ability to recover lost ground at the end of the race, making optimal positioning a non-trivial trade-off.
This setting naturally frames cross-country soaring as a collective exploration–exploitation problem, in which individual expertise competes with socially acquired information, much as in animal groups navigating uncertain environments~\cite{couzin2005effective,berdahl2018collective}. Targeted filtering of competition versus non-competition flights, or of isolated versus clustered flight conditions, could thus help rationalize, and possibly uncover, optimal strategies for group soaring.}

An appealing avenue for theoretical modeling would be to cast cross-country soaring as a \emph{foraging} problem in the spirit of first-passage and intermittent search theories: thermals act as ``resource patches'' where pilots ``refuel'' (gain altitude) before relocating. In the terminology of intermittent search theory~\cite{benichou2011intermittent}, transition phases correspond to fast \emph{non-reactive} phases primarily devoted to relocation during which thermal detection is negligible, whereas search phases correspond to slower \emph{reactive} phases during which thermals can be actively detected.
Within such a framework, one could aim to derive scaling predictions and/or optimality principles (e.g. maximizing net cross-country speed or minimizing mean first-passage times to the next usable thermal), and connect the empirically observed phase statistics to optimal intermittence results~\cite{benichou2011intermittent,benichou2005optimal,benichou2006twodim,redner2001firstpassage} (see also~\cite{almgren2015optimal}). In this interpretation, learning would primarily act on the {reactive} component of the dynamics, consistent with our phase-resolved results which suggest that the dominant learning signature is an improvement of the search strategy rather than of transition or climb kinematics.

More broadly, our results place human soaring flight alongside animal movement and active-matter systems in which superdiffusive transport emerges from the combined roles of persistence, intermittency, and atmospheric heterogeneity.

\section*{Acknowledgments}
We are grateful to Basile Dhote who contributed to the early stages of this work. We also thank Ortensia Forni, Clément Gouy-Paillier, Jorge Guirao Badiola, Roi Harel, Jutta Kurth, Iacopo Mastromatteo and Ran Nathan
 for fruitful discussions. This research was conducted within the Econophysics \& Complex Systems Research Chair, under the aegis of the Fondation du Risque, the Fondation de l’École polytechnique, the École polytechnique and Capital Fund Management. 
Portions of the writing and editing of this manuscript were assisted by ChatGPT, a large language model developed by OpenAI. All scientific content, analysis, and interpretations are the sole responsibility of the authors.

\section{Appendix}

\subsection{Flight time distributions}

\subsection{Flight phase segmentation}

To segment trajectories into transition, search, and climb phases, we use a Hidden Markov Model (HMM), following a standard state-space approach~\cite{rabiner2002tutorial}. The model is defined as a bivariate stochastic process $\{(Q_t,X_t)\}_{t\geq1}$, where $Q_t$ is a discrete-valued latent variable representing the flight phase at time $t$, and $X_t$ is an observable feature vector derived from the trajectory.
The latent process $\{Q_t\}$ is assumed to be a time-homogeneous Markov chain with three states (transition, search, climb). Conditional on $Q_t$, the observation vectors $\{X_t\}$ are assumed independent and identically distributed, with a state-dependent emission distribution. The latter encodes the statistical relationship between latent flight phases and the observed trajectory-derived features. In other words, it answers the question: if the system is in a given hidden state, what observations are likely to be produced?

\paragraph{Feature construction.}
At each time step, the observation vector is defined as
$X_t=\bigl(X_t^{v_z},\,X_t^{\mathrm{curv}},\,X_t^{\mathrm{str}}\bigr)$, 
where $X_t^{v_z}$ denotes the sign of the vertical velocity, $X_t^{\mathrm{curv}}$ quantifies curvature-angle sign persistence, and $X_t^{\mathrm{str}}$ is a binary indicator of trajectory straightness. All features are computed over a rolling time window of $30\,\mathrm{s}$.
Using binary variables avoids strong distributional assumptions and helps stabilize the emission statistics across heterogeneous landscapes, meteorological conditions, and pilot behaviors. The $30\,\mathrm{s}$ window corresponds approximately to the typical period of a full turn during thermalling, ensuring that the features capture phase-relevant dynamics while remaining insensitive to high-frequency fluctuations.

The vertical-speed feature $X_t^{v_z}$ is set to $1$ if the mean vertical speed over the window is positive and $0$ otherwise. Curvature-angle sign persistence $X_t^{\mathrm{curv}}$ is set to $1$ if the sign of the curvature angle is preserved for at least $90\%$ of samples within the window, and $0$ otherwise. Trajectory straightness $X_t^{\mathrm{str}}$ is set to $1$ if the absolute curvature angle remains below a prescribed threshold for at least $90\%$ of the window, and $0$ otherwise.
Note that because these features are computed over a rolling window, observations are not strictly independent at the original sampling frequency. However, conditional independence holds approximately at the $30\,\mathrm{s}$ scale, which is sufficient for identifying flight phases whose typical durations are on the order of minutes.

\begin{figure}[t!]
    \centering
    \includegraphics[width=\linewidth]{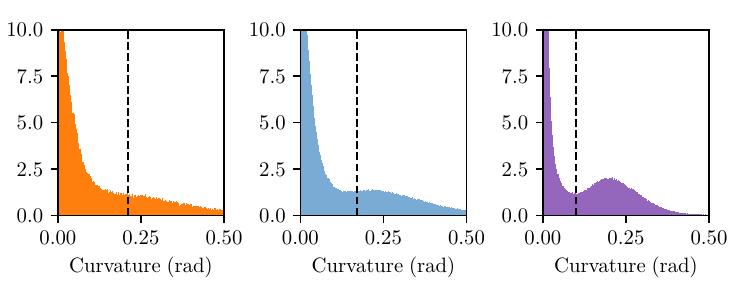}
    \caption{Distributions of absolute curvature angles in the $xy$-plane, for paragliders (left), hang gliders (center) and sailplanes (right). The straightness thresholds are indicated as vertical dashed lines.}
    \label{fig:threshold_straight}
\end{figure}

\paragraph{Curvature threshold selection.}
The straightness threshold is determined empirically from the distribution of absolute curvature angles in the horizontal plane (Fig.~\ref{fig:threshold_straight}). For sailplanes, the distribution exhibits a clear convex--concave--convex structure, which we interpret as reflecting distinct flight behaviors: nearly straight flight, thermalling or search turns with a characteristic curvature, and rare sharp turns. In this case, we select the threshold corresponding to the minimum of the first convex segment.
For paragliders and hang gliders, the curvature-angle distribution is monotonic, and the threshold is taken as the convex--concave inflection point. Thresholds are estimated using kernel density estimation (second-order for monotonic cases)~\cite{chen2017tutorial,sasaki2015direct}, yielding values of $0.21$, $0.17$, and $0.10$ radians for paragliders, hang gliders, and sailplanes, respectively.

\paragraph{Parameter estimation and decoding.}
The HMM parameters consist of the state transition matrix, the emission probabilities for the $2^3$ possible feature combinations, and the initial state distribution, for a total of $24$ parameters. These parameters are estimated by maximum likelihood using the Baum--Welch (EM) algorithm~\cite{baum1970maximization, rabiner2002tutorial}. Initial values are chosen to reflect the expected qualitative signatures of each phase, but the final parameter estimates are determined entirely by the data.
Given the estimated parameters, the most likely sequence of hidden states $(\hat Q_1,\ldots,\hat Q_T)$ is inferred using the Viterbi algorithm~\cite{viterbi2003error, forney2005viterbi}.

\bibliography{jeremie}

\end{document}